**Title:** Magnetic and Magnetocaloric Exploration of Fe rich (Mn,Fe)$_2$(P,Ge)

**Authors:** J. V. Leitão*[1], M. van der Haar[1], A. Lefering[1], E. Brück[1]

(1) Fundamental Aspects of Materials and Energy, Faculty of Applied Sciences, TU Delft, Mekelweg 15, 2629 JB Delft, Netherlands
* Corresponding author: J.C.Vieiraleitao@tudelft.nl

**Abstract:** We explored the Fe rich side of the (Mn,Fe)$_2$(P,Ge) magnetocaloric system.
The transition temperature of this system is extremely easy to tune with careful manipulation of Fe and Ge content as well as stoichiometrical proportions, which gives rise to the real possibility of lowering the price of this compound and thus make it economically viable for practical magnetocaloric applications.
Novel and unexpected magnetic properties observed in this system suggest an exciting potential for permanent magnet application in a limited concentration range.

## 1 – Introduction

Given the current technological trend towards energy efficiency and environmental friendliness, the magnetocaloric effect (MCE) is regarded as one of the best chances of creating a more efficient refrigeration system, alternative to the usual vapor compression technology. This effect, although known since the late 1920's [1-3], and its possible application in a magnetic cooling device since 1976 [4], is still to be effectively applied to a completely viable commercial room temperature cooling system. As such, one of the main focuses in MCE studies has been the constant search for more effective, cheap and non-toxic working materials that may solidly establish the possibility for such a working refrigerator. In this context, considerable attention has been paid to the use of transition metal alloys as a possible cheap and non-toxic material solution.

As a standard for the evaluation of the potential of a given material for magnetic cooling applications, the magnetic entropy change ($\Delta S$), derived from one of Maxwell's relations and expressed as Equation 1, is commonly used.

$$\Delta S(T)_{\Delta H} = \int_{Hi}^{Hf} \left( \frac{\partial M(T,H)}{\partial T} \right)_H dH \qquad (1)$$

Where T is temperature, M is magnetization and H is magnetic field.

As ΔS is given as the derivative of the magnetization as a function of temperature, it will be maximized around large jumps in magnetization, such as those present around the Curie Temperature (T$_C$) or other such magnetic phase transitions; also, mathematically speaking, it may be described as a function of the area between magnetic isotherms [5].

Among the recently studied systems for MCE the $(Mn,Fe)_2(P,Ge)$ system has shown great promise, having a highly tunable transition temperature and thermal hysteresis and high magnetization [6].

This system, however, has the great drawback of being considerably expensive due to the use of Germanium, which jeopardizes its use in a commercial magnetic refrigeration device.

Still, analysis of the results presented by Trung et al. [6], and more recently by Dung et al. [7] on the similar $(Mn,Fe)_{1.95}(P,Si)$ system, does reveal that increasing Fe composition in these systems triggers an increase in their transition temperature, which may then be compensated by decreasing Ge or Si content, respectively.

Obviously, among these two systems the $(Mn,Fe)_{1.95}(P,Si)$ stands out as the, by far, cheapest and safest, possessing remarkable magnetic and magnetocaloric properties. Unfortunately, the dramatic change in the *a* and *c* parameters observed in its crystal lattice during its first-order magneto-elastic transition implies that these samples may display a potentially crippling brittleness when they are thermally cycled [7], an added difficulty for their commercial application.

This has lead to the motivation of giving Fe rich $(Mn,Fe)_2(P,Ge)$ samples a better look, as the characteristics of this system's magneto-elastic transition suggests a more stable mechanical behavior, and the above mentioned results seem to point to the fact that Ge composition can be significantly reduced if Fe is increased, possibly making this system commercially viable.

Furthermore, this half of the $(Mn,Fe)_2(P,Ge)$ system's phase diagram appears to have been largely disregarded, as all recent publication appear to only focus on Mn rich samples, as shown in Figure 1. In that sense such an investigation will no doubt further enlighten not only this system but others belonging to the very promising family of $Fe_2P$ based compounds.

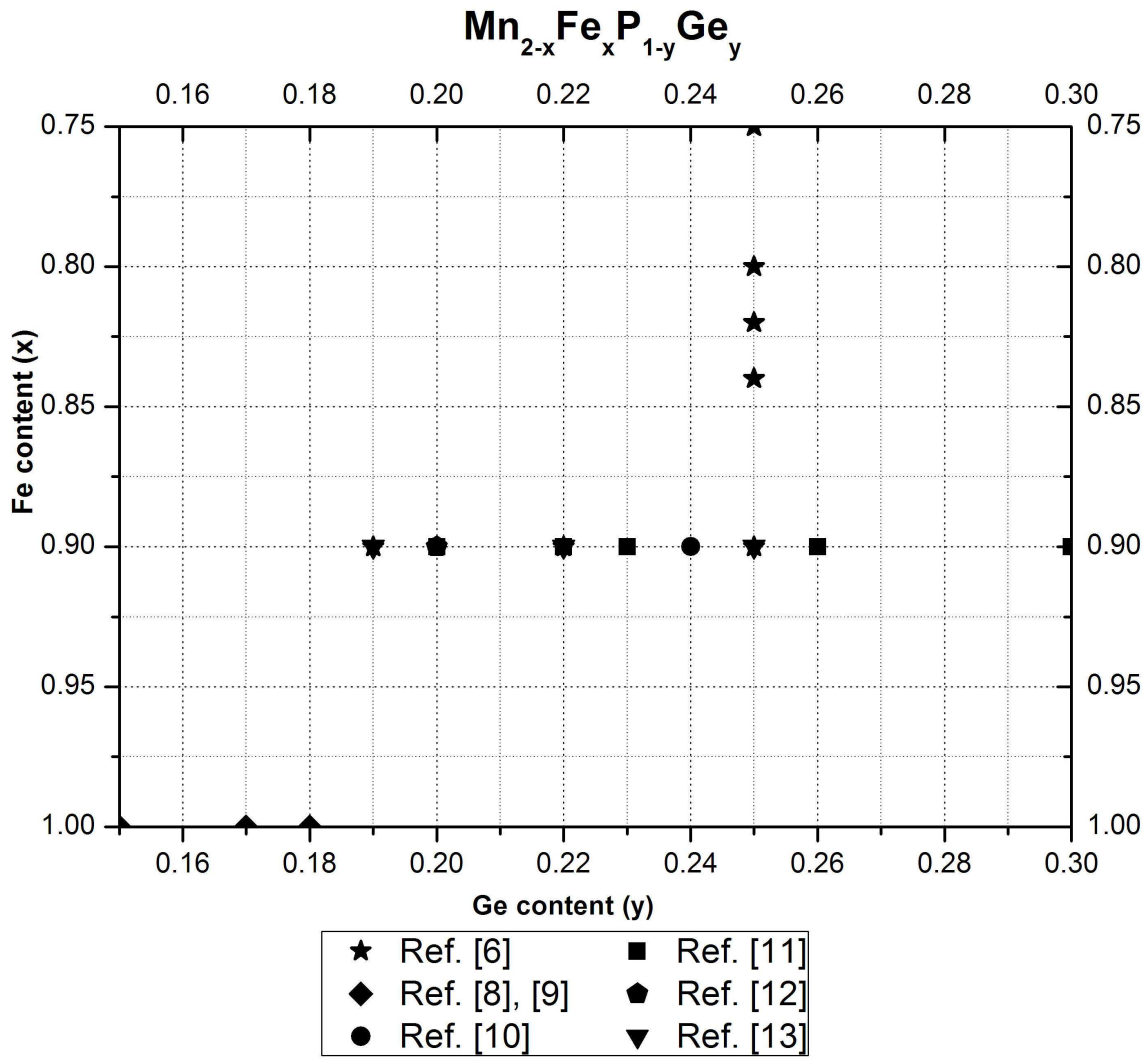

**Figure 1** – Studied compositions of recently published papers on the $(Mn,Fe)_2(P,Ge)$ system.

For this purpose various Fe rich samples belonging to the $(Mn,Fe)_2(P,Ge)$ system, and the closely related $(Mn,Fe)_{1.95}(P,Ge)$ system, have been produced so as their transition temperatures and overall magnetic properties could be monitored.

## 2 – Material overview

Both the $(Mn,Fe)_2(P,Ge)$ and the $(Mn,Fe)_{1.95}(P,Ge)$ systems crystallize in the $Fe_2P$-type hexagonal structure ($P\bar{6}2m$ space group) [6, 12]. The Fe and Mn transition metal atoms occupy the 3*f*-site at the tetrahedral (x1, 0, 0) position and the 3*g*-site at the pyramidal (x2, 0, 1/2) position. The non-metal P and Ge atoms can both occupy the 1*b*-site at the (0, 0, 1/2) position and in the 2*c*-site at the (1/3, 2/3, 0) position.

Considering the parent compound $Fe_2P$, as Mn atoms are added, in determent of Fe, these preferably occupy the 3g-sites, coplanar with the P and Ge atoms at the 1b-site, while the Fe atoms preferably occupy the 3f-sites, coplanar with the P and G atoms at the 2c-site

[14]. This property gives rise to what might be considered to be two basal planes in this structure, one containing Mn and the other Fe atoms, as depicted Figure 2.

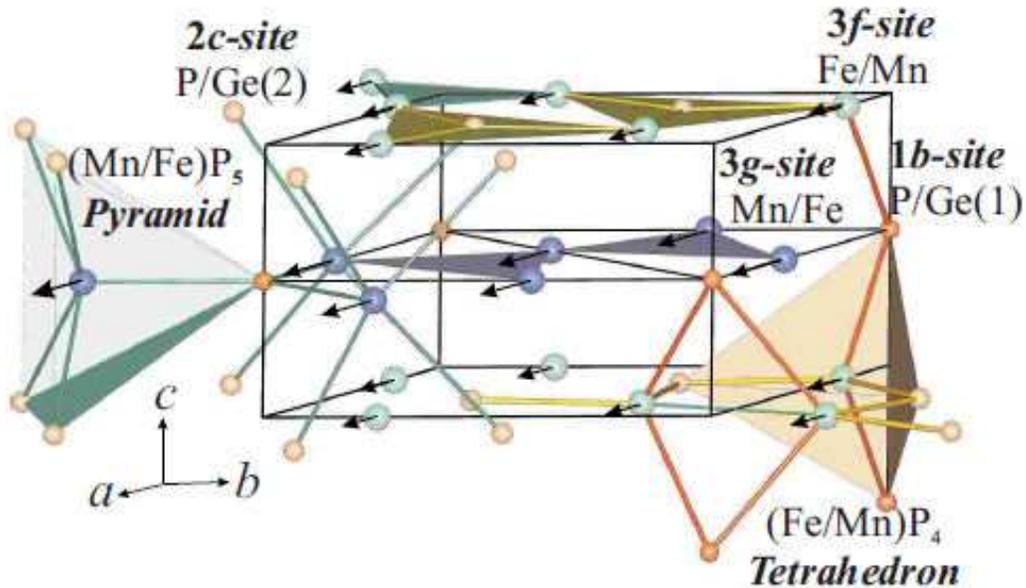

**Figure 2 - Crystal structure of the (Mn,Fe)$_2$(P,Ge) system presenting its several atomic sites; the arrows represent the magnetic moments of Mn and Fe atoms while the system is in its Ferromagnetic state [14].**

In terms of magnetism, this system is Paramagnetic at high temperatures and Ferromagnetic at low [14]. Its $T_C$ is extremely tunable by careful adjustment of both the Mn/Fe and the P/Ge ratio [6].

Accompanying the magnetic transition at $T_C$, this system also undergoes a contraction of the lattice parameter $c$ and an expansion of $a$ and $b$ [14], which can be identified in X-ray diffraction data as a shift in both the (300) and the (002) peak towards higher and lower angles respectively. This phenomenon is illustrated in Figure 3.

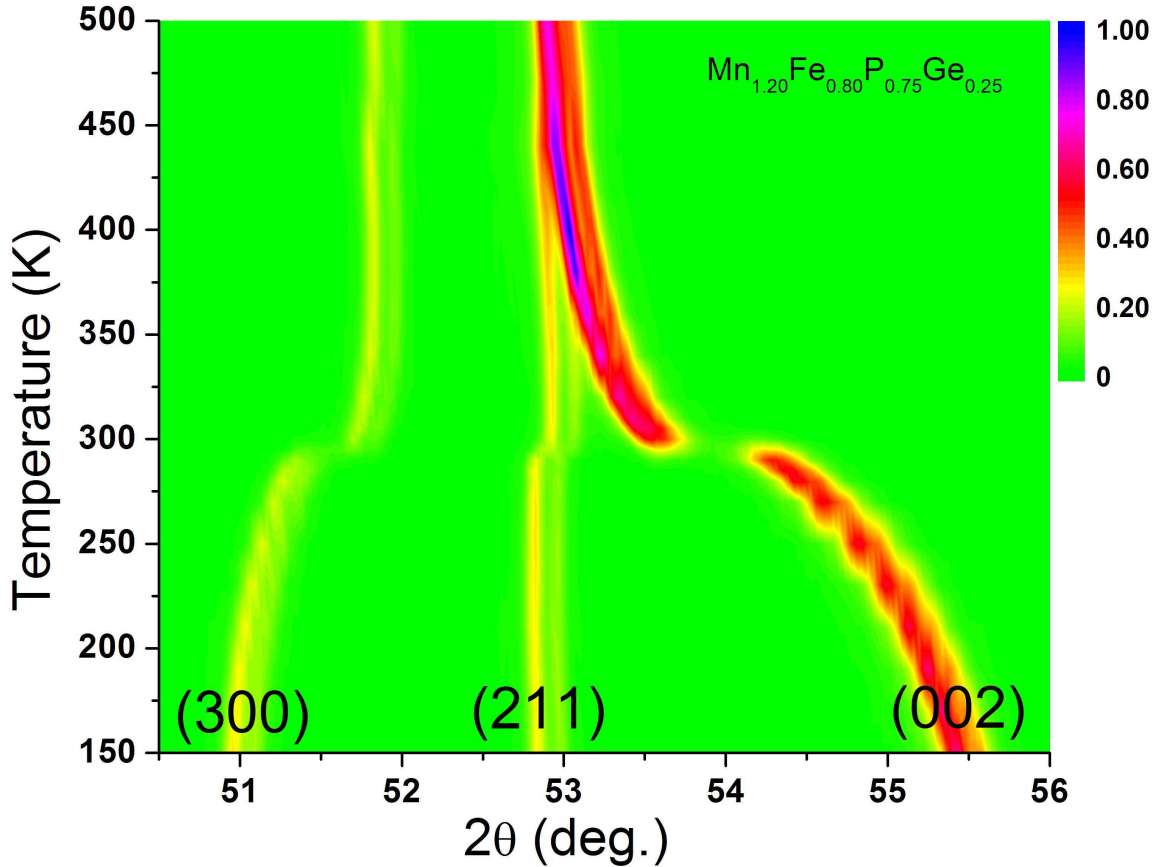

**Figure 3** - Contour plot of the angle change in the (300), (211) and (002) peaks versus temperature for $Mn_{1.2}Fe_{0.8}P_{0.75}Ge_{0.25}$, ilustrating the sharp discontinuity in lattice paramenters acompaning the magnetic transition [15].

## 3- Experimental Procedure and Characterization

3.1 – Sample Preparation

All samples produced belonging to the $(Mn,Fe)_2(P,Ge)$ and $(Mn,Fe)_{1.95}(P,Ge)$ systems were prepared from the appropriate amounts of 99+% iron powder, 99% red phosphorous powder, 99.5% binary $Fe_2P$ powder, 99.999% Ge chips and 99.9% manganese chips reduced at 600 ºC under a hydrogen atmosphere in order to remove oxides.
The samples were milled in a Fritsch Pulverisette planetary mill for 6 hours (3 hours with 5 minute breaks every 5 minutes to prevent overheating) at 360 rpm in 80 ml hardened steel crucibles, each containing fifteen 4 g hardened steel balls, amounting to a sample\ball ratio of 0.083(3) with the sample mass (5 g).
The samples were then compacted into 10 mm pellets with a pressure of 150 kgf/cm$^2$ and sealed into quartz tubes with an atmosphere of 200 mbar of argon. Finally these were annealed in a vertical resistive furnace at 1100 ºC for 10 hours, homogenized at 1000 ºC for 60 hours and then quenched to room temperature water.

3.2 – Characterization methods

In order to check the homogeneity and crystal structure of our samples, room temperature X-ray diffraction was performed in an X'Pert PRO X-ray diffractometer with Cu K$\alpha$ radiation from PANalytical. The resulting diffraction patterns were analyzed using the software X'Pert HightScore and FullProf's implementation of the Rietveld refinement method [16].

Magnetic measurements were performed in two different magnetometers, equipped with superconducting quantum interference devices (SQUID), a MPMS-5S and a MPMS XL model, both from Quantum Design.

The measurements taken were both temperature sweeps from 5 K to either 370 K (MPMS-5S) or 400 K (MPMS XL) with a fixed applied magnetic field, and field sweeps from -5 to 5 Tesla at fixed temperatures.

Further DSC (Differential scanning calorimetry) measurements were performed on those samples whose transition temperatures exceeded the temperature range of our magnetometers.

A Q2000 model from TA Instruments-Waters LLC was used for this end, performing temperature sweeps from 0 to 450 ºC at a rate of 20 ºC per minute.

**4 – Result and discussion**

4.1 – $(Mn,Fe)_2(P,Ge)$ system

Throughout the whole studied compositional range, all samples belonging to both the $(Mn,Fe)_2(P,Ge)$ and the $(Mn,Fe)_{1.95}(P,Ge)$ systems presented the same $Fe_2P$-type hexagonal structure, with the occasional occurrence of a minor MnO peak in their X-ray diffraction patterns. The lattice parameter *a* was found to increase with Ge addition and decrease with Fe, while the *c* parameter decreases with Ge and increases with Fe.

The $T_C$ of these systems was found to be quite easy to manipulate and tune with changes in both Fe and Ge content. As a preliminary study for the $(Mn,Fe)_2(P,Ge)$ system, the diagram in Figure 4 was assembled, enabling for a clear visualization and understanding of this tuning phenomenon.

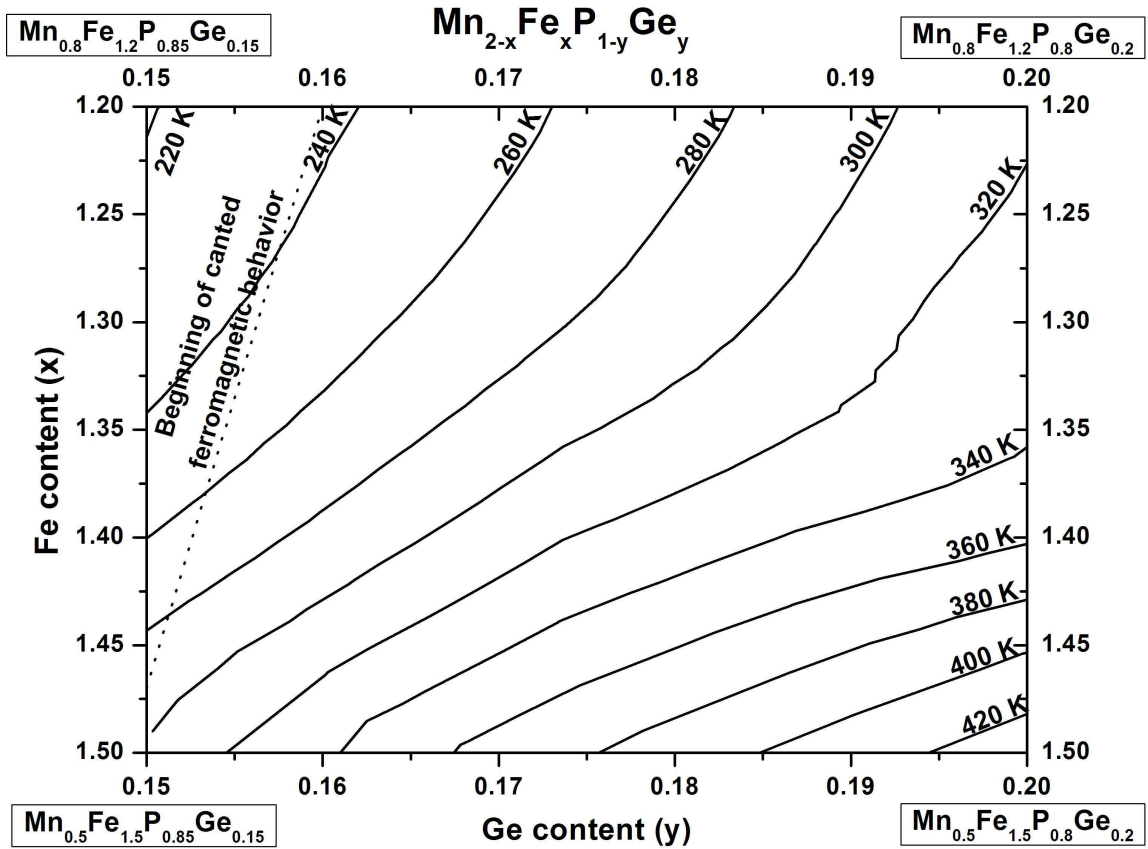

**Figure 4 - Evolution of T$_C$ with both Fe and Ge content in the (Mn,Fe)$_2$(P,Ge) system.**

The results displayed in Figure 4 show that in this compositional range T$_C$, while easily tunable, does not change linearly with neither Fe nor Ge concentration, contrarily with what had been previously reported by Trung et al. [6] and Brück at al.[17] for Mn rich (Mn,Fe)$_2$(P,Ge) compounds.

Also novel in this system is the detection of canted Ferromagnetic behavior at low Ge concentrations, which, if pushed to even lower Ge contents, turns full Antiferromagnetic, limiting the lowest usable Ge content in this system. This canted Ferromagnetic region seems to also be influenced by the Fe content, as increasing Fe pushes the occurrence of this behavior further into low Ge concentrations, eventually vanishing from our samples.

While, as can be seen in Figure 4, our initial hypothesis was correct, meaning that it is possible to reduce Ge by increasing Fe content and maintain T$_C$ around room temperature, this process has brought on an undesirable widening of the typically sharp transition between the Ferromagentic and the Paramagnetic state in this system, which in these concentrations no longer displays the characteristics of a first order phase transition.

Given the understanding that Equation 1 offers us regarding MCE, such an occurrence is extremely unfortunate for the prospect of applying Fe rich (Mn,Fe)$_2$(P,Ge) samples to any practical magnetic cooling device.

Figure 5 illustrates the above mentioned results.

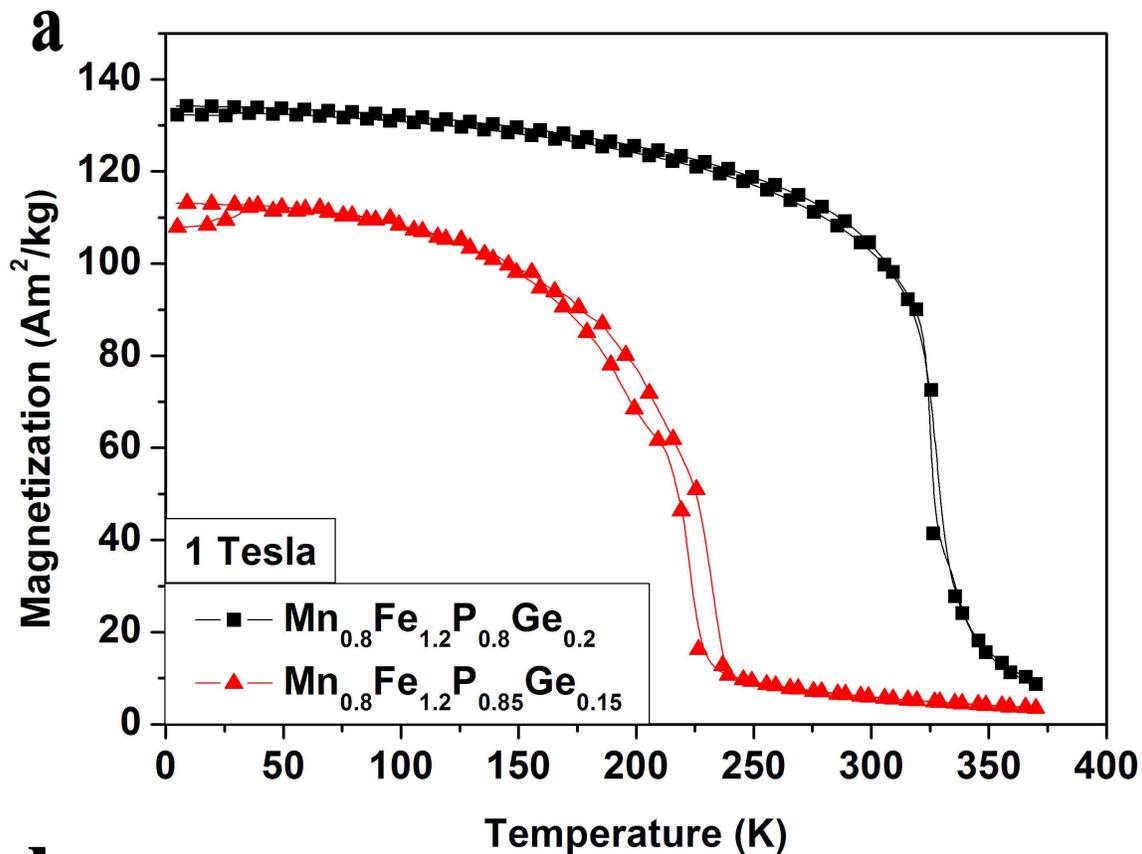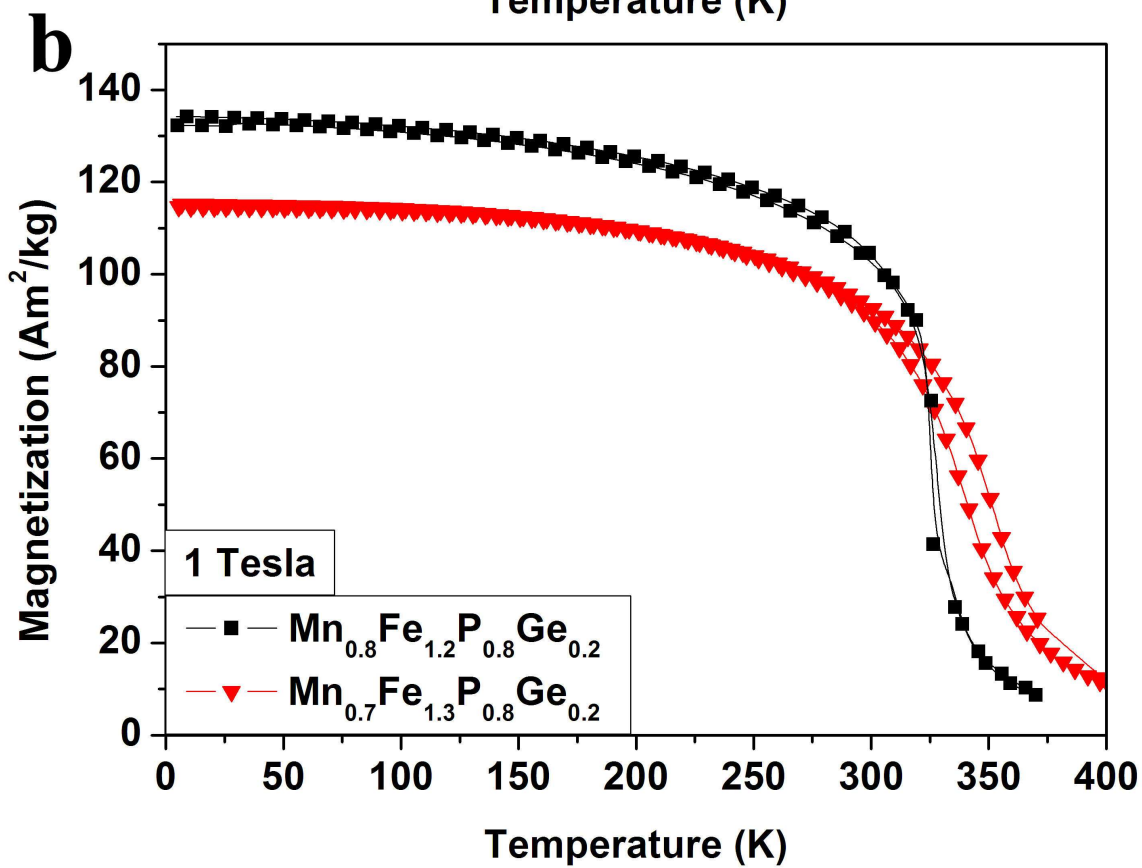

**Figure 5 - a)** Comparison between the Ferromagnetic and canted Ferromagnetic behavior in low Ge content samples in the (Mn, Fe)$_2$(P,Ge) system; **b)** Comparison between two samples with diferent Fe contents demonstrating the decrease in transition sharpenss promoted by the increase of Fe.

The loss of the first order behavior of the magneto-elastic transition also implies a disappearance in the discontinuity previously observed in the Mn rich magnetic and structural properties. This can be best observed by the monitoring of the (003) and (002) peaks in the X-Ray diffraction patterns of samples belonging to the Fe rich (Mn,Fe)$_2$(P,Ge).

While in the Mn rich side of the diagram we observe the already mentioned discontinuity in these two peaks around the transition temperature, exemplified in Figure 3, we now observe a very soft and slow movement in these both on them, as seen in Figure 6.

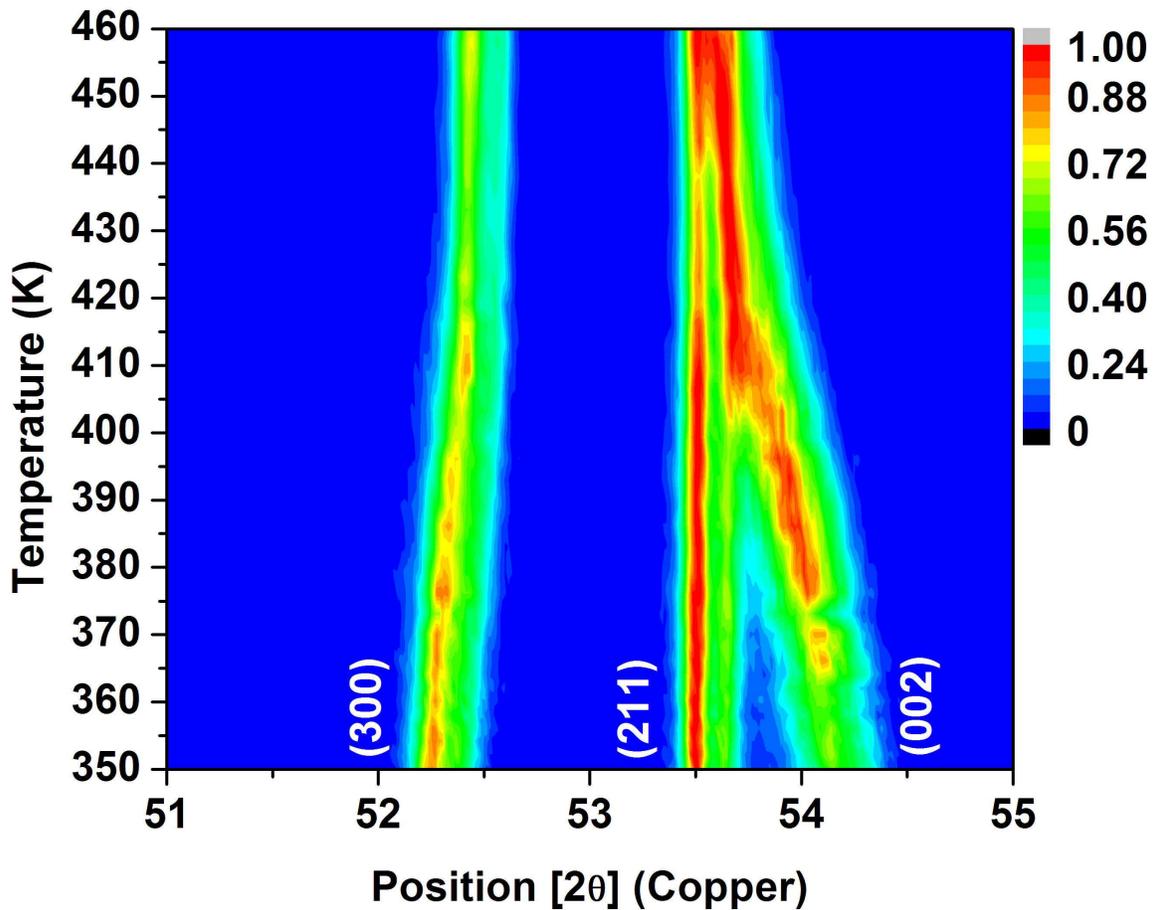

**Figure 6 -** Contour plot of the angle change in the (300), (211) and (002) peaks versus temperature for the Mn$_{0.6}$Fe$_{1.4}$P$_{0.8}$Ge$_{0.2}$ sample, demonstrating the slow and smooth character of the magneto-elastic transition. The transtion temperature of this sample has been determined as being 358 K by DSC measurments. The bar on the right of the figure represents the normalized peak intensity.

4.2 – (Mn,Fe)$_{1.95}$(P,Ge) system

The (Mn,Fe)$_{1.95}$(P,Ge) system on the other hand does not seem to be so negatively influenced by the increasing of Fe content, maintaining a usable sharp transition up to significantly high Fe concentrations.

This stability has enabled us, as was the original intent, to reduce Ge content down to Ge=0.12 (less than half of the value used by Trung et al. [6], for example). Lower Ge contents were found to trigger Antiferromagnetic behavior in this system, and as such we have assumed this to be the lowest possible Ge content in which this system still displays a sharp Ferro-Paramagnetic transition. See Figure 7.

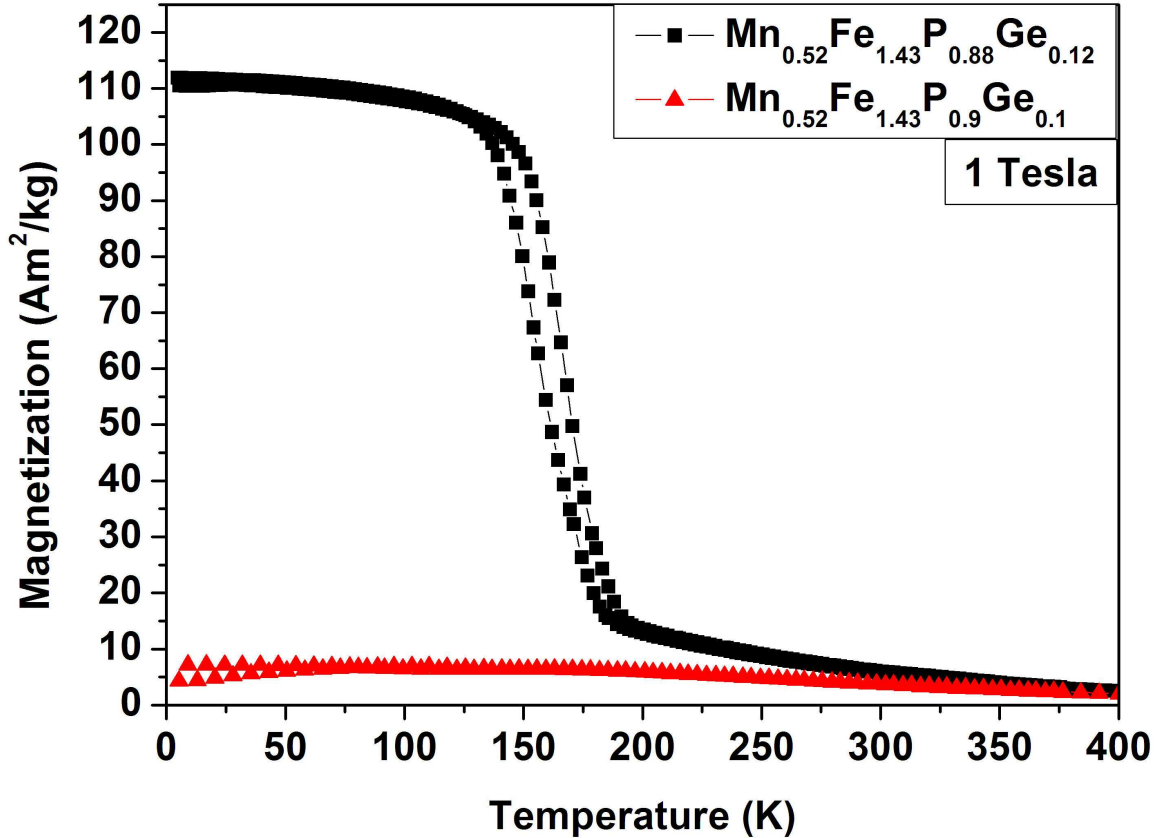

**Figure 7** - Comparison between the Ge=0.12 and Ge=0.1 magnetization versus temperature measurements in the (Mn,Fe)$_{1.95}$(P,Ge) system.

Having determined a usable Ge minimum, we were able to tune $T_C$ by changing Fe content, much in accordance with the behavior observed for the (Mn,Fe)$_2$(P,Ge) system and displayed in Figure 4. This, however, did still reveal itself challenging, as in this concentration range properties such as sharpness or magnetic behavior are extremely sensitive to small compositional changes. In this sense, while maintaining Ge=0.12, we have determined the Fe maximum (and consequently the maximum $T_C$) in which this system still displays usable characteristics for a magnetic cooling device. See Figure 8.

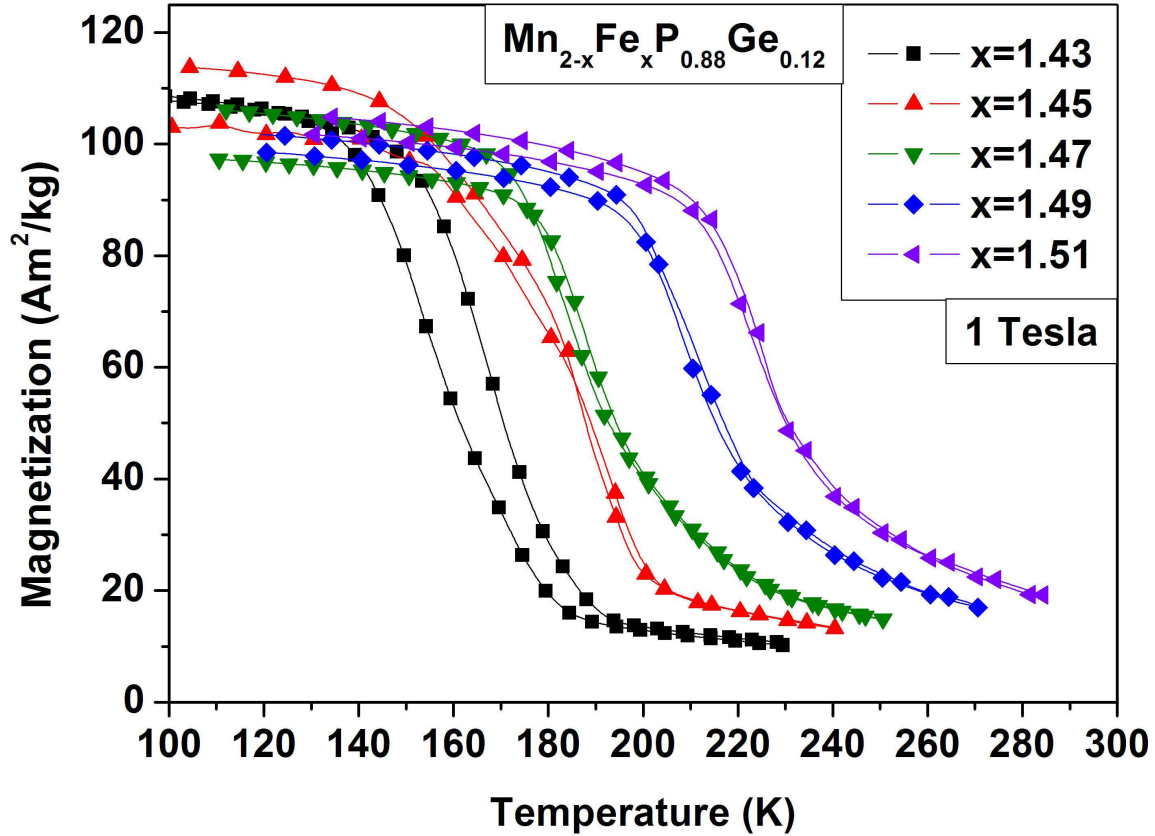

**Figure 8** - Magnetization versus temperature curves for $Mn_{2-x}Fe_xP_{0.88}Ge_{0.12}$ samples for an applied magnetic field of 1 Tesla, demonstrating the change in $T_C$ with increasing Fe content.

We have thus determined a maximum $T_C$ of about 225 K for a maximum Fe content of 1.51, within the determined Ge minimum (0.12).

Although these results may be seen less ideal than the previously mentioned ones from Trung et al. [6], such is the trade off of cheaper samples with improved mechanical properties, which none the less still possess many possible applications in low temperature applications or in magnetocaloric material cascading in practical cooling devices.

4.2.1 – Permanent magnet potential

Surprisingly, during the course of this study certain limited compositional areas of Mn and Ge poor $(Mn,Fe)_{1.95}(P,Ge)$ have been found to possess very exciting magnetic properties, which may indicate a definite potential and impact for future permanent magnet applications.

In this perspective, the relevant intrinsic properties a material should display for permanent magnet applications can be listed as such: 1) high saturation magnetization; 2) magneto-crystalline anisotropy and uniaxial crystal structure; 3) high $T_C$ [18].

Magnetization versus temperature measurements performed on the sample $Mn_{0.1}Fe_{1.85}P_{0.9}Ge_{0.1}$ suggests such properties as described above. See Figure 9.

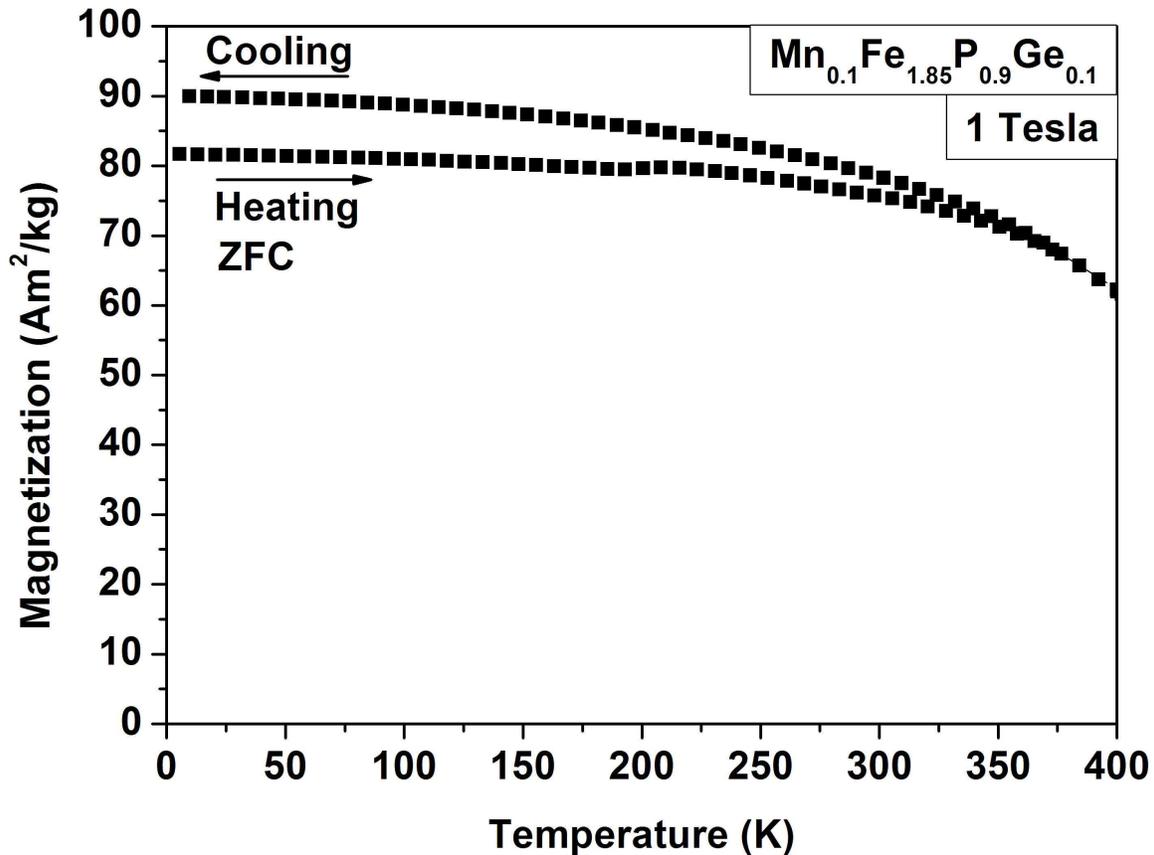

Figure 9 - Magnetization versus temperature for the $Mn_{0.1}Fe_{1.85}P_{0.9}Ge_{0.1}$ sample, revealing a significant magnetic anisotropy due to the effects of thermal motion in the sample's magnetic moments.

Besides the significant magnetic moment measured, it can be clearly observed that there is a substantial difference in magnetic behavior upon heating and cooling. This can be explained by the fact that this sample was zero field cooled to 5 K.

At low temperature, due to a lack of thermal motion, the sample's magnetic moments are unable to effectively align with the applied magnetic field. As the sample is subsequently warmed, thermal motion increases and the moments are able to align. When the sample is once again cooled the thermal motion once again decreases but the moments that have already aligned with the field will remain so, giving rise to the difference in the heating and cooling curves seen in Figure 9. This difference may then be taken as a strong indication for the presence of significant magnetic anisotropy in this sample.

In order to further study this phenomenon, two different types of magnetization versus field measurements were performed. The promising samples were reduced to a fine powder and, firstly, mixed with varnish, being allowed to solidify with a random grain orientation. Secondly, the same process was followed but the varnish solidified while the powder was under a magnetic field, meaning that the grains were able to align with the field and in this way simulate something similar to a single crystal. Figure 10 displays the results from these measurements.

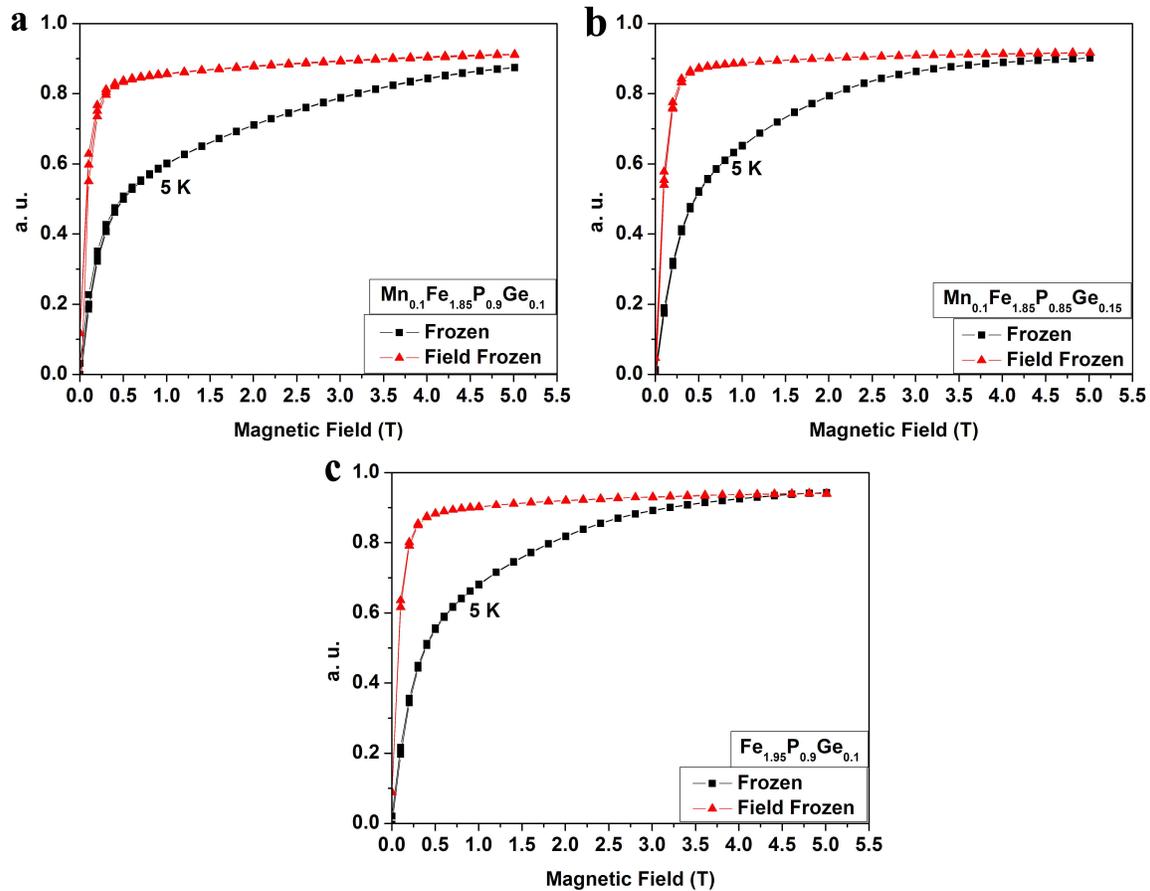

**Figure 10** - a) Magnetization versus applied magnetic field at 5 K for the $Mn_{0.1}Fe_{1.85}P_{0.9}Ge_{0.1}$ sample. As can be observed there is a clear difference in magnetic behavior between the field aligned and the randomly alighted sample, indicating the presence of strong magnetic anisotropy; b) Magnetization versus applied magnetic field at 5 K for the $Mn_{0.1}Fe_{1.85}P_{0.85}Ge_{0.15}$ sample, demonstrating a similar, but not as pronounced, magnetic anisotropy as the $Mn_{0.1}Fe_{1.85}P_{0.9}Ge_{0.1}$ sample; c) Magnetization versus applied magnetic field at 5 K for the $Fe_{1.95}P_{0.9}Ge_{0.1}$ sample, also demonstrating the same clear difference in magnetic behavior but giving a good insight on the influence of the small Mn content present on the $Mn_{0.1}Fe_{1.85}P_{0.9}Ge_{0.1}$ sample

The clear difference in magnetic behavior between the randomly frozen and the field frozen sample suggests a strong anisotropy, associated with an easy axis that allows for a rapid magnetic saturation.

Analyzing the results depicted in Figure 10 it can be concluded that indeed Ge is a fundamental element in the occurrence of this behavior, although its ideal content does seem to be a complex issue to ascertain. Comparing between the $Mn_{0.1}Fe_{1.85}P_{0.9}Ge_{0.1}$ and the $Mn_{0.1}Fe_{1.85}P_{0.85}Ge_{0.15}$ samples, we notice a reduction in magnetic anisotropy in the later, while completely removing Ge appears to simply remove the occurrence of this behavior [19], as it gives room for other already studied states of metamagnetism and antiferromagnetism related to the presence of small amounts of Mn in this system [20].

The small amount of Mn in these samples does not seem to be negligible either, as it does also seem to play some part in this behavior given that the $Fe_{1.95}P_{0.9}Ge_{0.1}$ sample, while still possessing a clear magnetic behavior difference, reaches its saturation magnetization much more rapidly in its randomly frozen measurement than all other samples. This indicates that indeed the small amount of Mn present in the other two samples does contribute to increase the large anisotropy detected.

The evidence of anisotropy was also verified by X-ray diffraction. Measurements were conducted in regular randomly aligned powder and in a magnetic field aligned powder, so as to once again simulate a single crystal. See Figure 11.

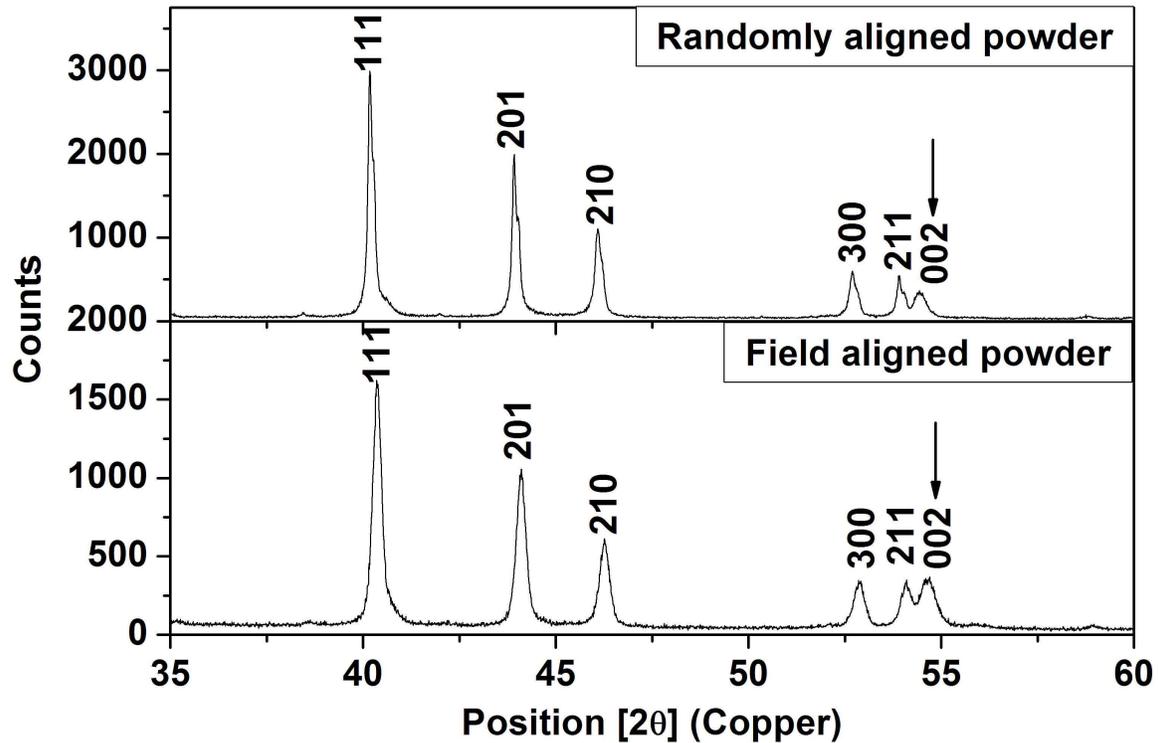

Figure 11 - X-Ray diffraction measurements of $Mn_{0.1}Fe_{1.95}P_{0.85}Ge_{0.15}$. The top graph shows a regular randomly aligned powder and the bottom one a field aligned powder, in which it is possible to observe an increase in the intensity of the (002) reflection, indicating an alignment along the c-direction.

The results from these measurements, while showing a typical $Fe_2P$ hexagonal structure, also reveal a clear increase in the (002) peak intensity relatively to the other peaks, which indicates the presence of an easy axis along the c-direction, once again underlying the presence of magnetic anisotropy in this sample.

The only draw backs of this surprising discovery so far are the obvious lack of coercivity, or a broad hysteresis loop, and the still relatively low $T_C$ presented by the $Mn_{0.1}Fe_{1.85}P_{0.9}Ge_{0.1}$ sample (from the studied batch the one with the best magnetic properties) of about 430 K. This value, although much higher than that of pure $Fe_2P$ [21], is still too low for a practical permanent magnet application.

It should also be noted that the use of Ge in this system further jeopardizes its practical use in large scale permanent magnet applications [22], but none the less one cannot stress enough that this was indeed the first time that such a potential has been observed in an $Fe_2P$-type system. Such an observation cannot be understated, and it may now open new and exciting opportunities for novel permanent magnet alloys in this rich family of materials. Further study and research is demanded.

## 5 – Conclusion

Given their attractive mechanical properties, a study on the $(Mn,Fe)_2(P,Ge)$ and $(Mn,Fe)_{1.95}(P,Ge)$ systems was conducted with the purpose of making such systems economically viable for magnetic cooling applications.

The mapping of a limited compositional range of the $(Mn,Fe)_2(P,Ge)$ system revealed a strong non-linear $T_C$ dependence with both the Fe and Ge content, giving rise to the possibility of the reduction of the expensive Ge by the increase of Fe content. Such map, beyond the immediate application in the study of Mn-Fe-P-Ge systems, also offers an understanding on the general behavior of $Fe_2P$-type systems at high Fe concentrations.

The magnetic behavior of Fe rich $(Mn,Fe)_2(P,Ge)$ revealed to be quite complex, with the discovery of canted Ferromagnetic behavior at low Ge concentrations. Unexpectedly, the increase in Fe content also gives rise to a decrease in the Ferro-Paramagnetic transition sharpness, symptomatic with the disappearance of the first order behavior in this transition, an unfortunate characteristic that makes this system unviable for MCE applications.

The $(Mn,Fe)_{1.95}(P,Ge)$ system maintains a sharp transition with a high magnetic moment at high Fe concentrations, with the possibility of lowering Ge content down to 0.12, effectively reducing the cost of this system to under half of the originally studied compositions. Given the difficult balance of magnetic properties and element concentrations, $T_C$ cannot, however, be raised above 225 K, roughly, but it is none the less easily tuned below this temperature.

Remarkable and exciting properties were detected in a limited concentration range of Mn and Ge poor $(Mn,Fe)_{1.95}(P,Ge)$, which may indicate a tangible possibility of using $Fe_2P$-type alloys for permanent magnet applications. This result is still not final but it points to a new and exciting field of permanent magnet research.

## Acknowledgments

The authors wish to acknowledge the funding from BASF Future Business and FOM (Stichting voor Fundamenteel Onderzoek der Materie), under the Industrial Partnership Programme IPP I18 of the 'Stichting voor Fundamenteel Onderzoek der Materie (FOM)' which is financially supported by the 'Nederlandse Organisatie voor Wetenschappelijk Onderzoek (NWO)'.## References

[1] P. Debye, Ann. Phys., 81 (1926) 1154;
[2] W. F. Giauque, J. Am. Chem. Soc., 49 (1927) 1864;
[3] W. F. Giauque and D.P. MacDougall, Phys. Rev., 43 (1933) 768;
[4] K. A. Gschneidner Jr. and V. K. Pecharsky, Int. Jour. of Refri., 31 (2008) 945;
[5] A.M. Tishin and Y.I. Spichkin, *The Magnetocaloric Effect and Its Applications*, Bodmin, MPG Books Ltd, 2003.
[6] N. T. Trung, Z. Q. Ou, T. J. Gortenmulder, O. Tegus, K. H. J. Buschow, and E. Brück, Appl. Phys. Lett., 94, 102513 (2009);
[7] N. H. Dung, L. Zhang, Z. Q. Ou, and E. Brück, Appl. Phys. Lett., 99, 092511 (2011);